\documentclass[twocolumn,showpacs,preprintnumbers,elsart]{revtex4}
\usepackage{eurosym}
\usepackage{makeidx}
\usepackage{amssymb}
\usepackage{amsmath}
\usepackage{mathrsfs}
\usepackage{graphicx}
\usepackage{dcolumn}
\usepackage{bm}
\usepackage[center]{subfigure}
\usepackage{color}
\usepackage{indentfirst}
\usepackage{booktabs}

\makeatletter

\newcommand{\Rmnum}[1]{\expandafter\@slowromancap\romannumeral #1@}
\makeatother

\begin{document}

\title{Excited states of two-dimensional solitons supported by the
spin-orbit coupling and field-induced dipole-dipole repulsion}
\author{Chunqing Huang$^{1}$, Yuebo Ye$^{1}$, Shimei Liu$^{1}$, Hexiang He$%
^{1}$, Wei Pang$^{2}$, Boris A. Malomed$^{3,4,1}$, and Yongyao Li$^{1}$}
\email{yongyaoli@gmail.com}
\affiliation{$^{1}$School of Physics and Optoelectronic Engineering, Foshan University,
Foshan $528000$, China\\
$^{2}$Department of Experiment Teaching, Guangdong University of Technology,
Guangzhou 510006, China\\
$^{3}$Department of Physical Electronics, School of Electrical Engineering,
Faculty of Engineering, and the Center for Light-Matter Interaction, Tel
Aviv University, Tel Aviv 69978, Israel\\
$^{4}$ITMO University, St. Petersburg 197101, Russia}

\begin{abstract}
It was recently found that excited states of semi-vortex and mixed-mode
solitons are unstable in spin-orbit-coupled Bose-Einstein condensates (BECs)
with contact interactions. We demonstrate a possibility to stabilize such
excited states in a setting based on repulsive dipole-dipole interactions
induced by a polarizing field, oriented perpendicular to the plane in which
the dipolar BEC is trapped. The strength of the field is assumed to grow in
the radial direction $\sim $ $r^{4}$. Excited states of semi-vortex solitons
have vorticities $S$ and $S+1$ in their two components, each being an
eigenstate of the angular momentum. They are fully stable up to $S=5$.
Excited state of mixed-mode solitons feature interweaving necklace
structures with opposite fractional values of the angular momentum in the
two components. They are stable if they are built of dominant angular
harmonics $\pm S$, with $S\leq 4$. Characteristics and stability of these
two types of previously unknown higher-order solitons are systematically
analyzed. Their characteristic size is $\sim 10$ $\mathrm{\mu }$m, with the
number of atoms $\lesssim 10^{5}$.\\
\end{abstract}

\pacs{03.75.Lm; 42.65.Tg; 47.20.Ky; 05.45.Yv}
\maketitle

%\email{chunqinghuang@qq.com}

%$^{3}$ Department of Physical Electronics, School of Electrical Engineering, Faculty of Engineering, Tel Aviv University, Tel Aviv $69978$, Israel}

%\email{yongyaoli@gmail.com}
% Force line breaks with \\
% \textcolor{red}{[]}

\section{Introduction and the model}

Stabilizing bright solitary waves and vortices in the two- and
three-dimensional (2D and 3D) free space with cubic nonlinearity remains a
problem of great interest in nonlinear optics and studies of Bose-Einstein
condensates (BECs), as well as in other areas \cite{review,Dumitru}. A
well-known challenging problem is that the ubiquitous cubic local attractive
nonlinearity makes multidimensional solitons unstable against collapse.
Diverse methods have been proposed to suppress this instability. In
particular, in nonlinear optics stable 2D optical solitons have been
predicted and created in media with saturable \cite{Segev1994}, quadratic
\cite{Mihalache2006}, cubic-quintic \cite{Mihalache22006}, and nonlocal
nonlinearities \cite{Mihalache32006,Skupin,Rotschild2006}, which do not
cause collapse.

It has been predicted\ too that long-range dipole-dipole interactions can
create 2D matter-wave solitons in BECs with permanent atomic/molecular
magnetic or electric dipole moments \cite%
{Pedri2005,Nath2008,Tikhonenkov2008,Tikhonenkov22008,Raghunandan2015,
Jiasheng}. Because dipole-dipole interactions can be tuned to
be isotropic or anisotropic by choosing the orientation of the external
polarizing field with respect to the system's plane, this makes the dipolar
BECs appropriate media for simulating 2D and 3D solitons and solitary
vortices.

In addition to the stabilization induced by dipole-dipole interactions, it
has also been predicted that 2D and 3D solitons can be stabilized in spinor
(two-component) BECs with the help of Rashba-type spin-orbit (SO) coupling
\cite{Wilson2013,SVS1,SVS2,SVS3,SVS4,Guihua2017,Yongchang,QD2017,Bingjin2017}%
. Similar to dipole-dipole interactions, SO coupling in BECs \cite%
{nature09887,zhaihui2015,Yongping2016} provides broad tunability for the
formation of solitons \cite%
{Sakaguchi2,Kartashov1D,Kartashov2017,Kartashov2D,Xuyoong2015,YXU2013,Salasnich2014}%
. To date, two types of stable multidimensional solitons, \textit{viz}.,
semi-vortices (SVs, also called half-vortices \cite{Drummond}) and
mixed-mode (MM) states, have been predicted in SO-coupled BECs. More complex
self-trapped modes, which may be considered as excited states of SVs and
MMs, produced by adding the same vorticity to both components of the binary
soliton, are also supported by the Rashba SO coupling. Excited states of SVs
and MMs exhibit complex patterns, which may be of considerable interest to
the soliton physics, if they can be made stable. In reality, all the
previously studied excited states were found to be unstable under the action
of local and nonlocal attractive nonlinearities \cite{SVS1,QD2017}, quickly
or gradually decaying back to their ground-state counterparts, or completely
losing the soliton structure.

The objective of this work is to predict stable excited states and study
their properties in SO-coupled BECs, using a nonlocal \emph{repulsive}
nonlinearity. Recently, we have found that BECs with repulsive dipole-dipole
interactions can support stable 2D gap solitons with large values of
embedded vorticity, shaped as vortex rings \cite{Huang2017}. This finding
suggests that repulsive dipole-dipole interactions may be an appropriate
means for stabilizing other types of 2D solitons, including excited states
in SO-coupled BECs.

Of course, repulsive nonlinearity cannot create bright soliton in free space
(without the help of a trapping potential; in models of the nonlinear
Dirac/Weyl type, which do not contain the usual kinetic-energy terms,
repulsive dipole-dipole interactions can support bright gap soliton in the
free space for the SO-coupled BEC, under the action of the Zeeman splitting
\cite{SOCgapsoliton}). On the other hand, it was reported that bright
fundamental and vortex 2D solitons can be supported by spatially-patterned
repulsive dipole-dipole interactions, induced by a nonuniform polarizing
field, the strength of which grows from the center to periphery, as a
function of distance $r$, at any rate faster than $r^{3}$ \cite%
{Yongyao2013,Abdullaev2014}. Similar settings, using spatially varying
contact (local) self-repulsion, the strength of which grows, in the space of
dimension $D$, faster than $r^{D}$ \cite{Borovkova2011}-\cite{Driben2014},
have been reported to support robust families of complex modes, such as
hopfions \cite{localgrow}, which carry two independent topological charges.

We aim to predict stable excited states of SVs and MM in a similar 2D
system, using the repulsive interaction whose strength growth from the
center to periphery. While this possibility can be realized with both local
(contact) and nonlocal (dipole-dipole) interactions, it is easier to produce
stable excited states, with larger values of the added vorticity, in the
latter case. Therefore, we here consider long-range dipole-dipole
interactions, adopting the scheme introduced in Ref. \cite{Yongyao2013} (the
case of contact repulsive interactions, which produces essentially different
results, will be considered elsewhere \cite{Rongxuan}): an effectively two-dimensional BEC
composed of atoms carrying electric dipole moments, $g(r)$, which are
induced by a polarizing field, $E(r)$, directed perpendicular to the
system's $(x,y)$ plane, with the field's strength growing along the radial
coordinate, $r$:%
\begin{equation}
g(r)=\chi E(r),  \label{E}
\end{equation}%
where $\chi $ is the atomic polarizability. The spatially modulated field
may be imposed by an external capacitor with the separation between its
electrodes decreasing with the increase of $r$ \cite{Yongyao2013}. Because
two components of the spinor BEC corresponds to different hyperfine states
of the same atom, identical dipole moments are induced in both components.
In the mean-field approximation, the dynamics of the spinor wave function, $%
\psi =(\psi _{+},\psi _{-})$, is governed by the scaled form of the
Gross-Pitaevskii equation:
\begin{eqnarray}
&&i\partial _{t}\psi _{\pm }=-{\frac{1}{2}}\nabla ^{2}\psi _{\pm }\pm
\lambda \hat{D}^{[\mp ]}\psi _{\mp }+g(\mathbf{r})\psi _{\pm }\times   \notag
\\
&&\int d\mathbf{r^{\prime }}R(\mathbf{r}-\mathbf{r^{\prime }})g(\mathbf{%
r^{\prime }})(|\psi _{+}(\mathbf{r^{\prime }})|^{2}+|\psi _{-}(\mathbf{%
r^{\prime }})|^{2}),  \label{fulleq}
\end{eqnarray}%
where $\hat{D}^{[\pm ]}=\partial _{x}\pm i\partial _{y}$ are the SO-coupling
operators with strength $\lambda $. Because the dipoles are perpendicular to
the 2D plane, the kernel of the dipole-dipole interactions is
\begin{equation}
R(\mathbf{r}-\mathbf{r^{\prime }})={1/(\epsilon ^{2}+|(\mathbf{r}-\mathbf{%
r^{\prime }})|^{2})^{3/2}},  \label{kernel}
\end{equation}%
where cutoff $\epsilon $ is determined by the confinement length $a_{\perp }$
in the transverse dimension, whose typical size in underlying physical units
is
\begin{equation}
a_{\perp }\sim 3~\mathrm{\mu m}  \label{epsilon}
\end{equation}%
\cite{Stoof}. As mentioned above, it was demonstrated in Ref. \cite%
{Yongyao2013} that the spatially modulated repulsive dipole-dipole
interactions can create 2D solitons, provided that the magnitude of the
locally induced dipole moment grows in $r$ faster than $r^{3}$, therefore we
here adopt the modulation profile
\begin{equation}
g(r)=\alpha r^{4}+g_{0},  \label{g0}
\end{equation}%
with $\alpha >0$ and $g_{0}\geq 0$.

Stationary solutions to Eq. (\ref{fulleq}) are sought for in the usual form,
$\psi _{\pm }(\mathbf{r},t)=\phi _{\pm }(\mathbf{r})e^{-i\mu t}$, where $%
\phi _{\pm }$ are stationary wave functions and $\mu $ is a real chemical
potential. Solitons are characterized by the total norm, which is
proportional to the number of atoms in the binary BEC:
\begin{equation}
N=N_{+}+N_{-}=\int d\mathbf{r}(|\phi _{+}|^{2}+|\phi _{-}|^{2}).  \label{N}
\end{equation}%
The system's energy is
\begin{equation}
E=E_{\mathrm{K}}+E_{\mathrm{DD}}+E_{\mathrm{SO}},  \label{Energy}
\end{equation}%
where $E_{\mathrm{K}}$, $E_{\mathrm{DD}}$ and $E_{\mathrm{SO}}$ are the
kinetic, dipole-dipole, and SO-coupling energies, respectively:
\begin{equation}
\begin{aligned} &E_{\mathrm{K}} ={1\over 2}\int
d\mathbf{r}\left(|\nabla\phi_{+}|^{2}+|\nabla\phi_{-}|^{2}\right),\\
&E_{\mathrm{DD}}={1\over2}\iint
d\mathbf{r}d\mathbf{r'}g(\mathbf{r})\left(|\phi_{+}(\mathbf{r})|^2+|%
\phi_{-}(\mathbf{r})|^2\right)\times\\ &\quad\quad
R(\mathbf{r}-\mathbf{r'})g(\mathbf{r'})(|\phi_{+}(\mathbf{r'})|^{2}+|%
\phi_{-}(\mathbf{r'})|^{2}),\\ &E_{\mathrm{SO}}=\lambda \int
d\mathbf{r}(\phi^{\ast}_{+}\hat{D}^{[+]}\phi_{-}-\phi^{\ast}_{-}%
\hat{D}^{[-]}\phi_{+}). \end{aligned}  \label{threeenergy}
\end{equation}

While all the quantities in Eqs. (\ref{fulleq})-(\ref{threeenergy}) are
written in the scaled form, therefore units are not necessary in figures
displayed below, it is relevant to summarize here estimates for the relevant
quantities in physical units, using, in particular, estimates elaborated in
Refs. \cite{Drummond} and \cite{Yongyao2013} for related settings. First,
characteristic values of the atomic or molecular electric polarizability
relevant to experiments with ultracold gases may be taken as $\sim 100{%
\mathrm{\mathring{A}}}^{3}$, with the corresponding atomic/molecular weight
being $M\sim 100$ \cite{chi}. The effective scattering length of the
dipole-dipole interactions, which is sufficient for the formation of
localized modes is%
\begin{equation}
a_{\mathrm{DD}}\sim 1~\mathrm{nm}  \label{aDD}
\end{equation}%
\textrm{\ }\cite{Pfau}. As it follows from Eqs. (\ref{E})-(\ref{kernel}),
the corresponding intensity of the dipole-dipole interactions is induced by
the polarizing dc electric field in a range of $\sim 10$ kV/cm (which means
that voltage $\sim 10$ V should be applied to the polarizing capacitor with
the separation $\sim $ $10$ $\mathrm{\mu }$m between its electrodes). The
results reported below are relevant to the experimental realization if, in
the scaled units adopted here, $x=1$ corresponds to the physical distance $%
\sim 10$ $\mathrm{\mu }$m. This, in turn, implies that normalization $%
\lambda =1$ adopted below in the scaled units represents the physical
strength of the SO coupling $\lambda \sim 10^{-6}$ g$\cdot \mathrm{\mu }$m$%
^{3}$/(ms)$^{2}$. More appropriate, in this context, is the estimate for the
effective length of the SO coupling,
\begin{equation}
a_{\mathrm{SO}}=\hbar ^{2}/\left( m\lambda \right) \sim 0.5~\mathrm{\mu m},
\label{aSO}
\end{equation}%
where $m$ is the atomic mass in physical units. This range of values of the
SO-coupling strength is accessible to the current experiments \cite%
{nature09887,zhaihui2015,Yongping2016}, \cite{Drummond}. Finally, it follows
from estimates given by Eqs. (\ref{epsilon}), (\ref{aDD}) and (\ref{aSO})
that values of scaled norm $N$ in the range of $1\sim 5$, which appear in
the results reported below, correspond to the number of atoms $10^{4}\sim
10^{5}$ in the quasi-2D soliton, which should make the observation of the so
predicted self-trapped states definitely possible \cite{Hulet}.

\section{Semi-vortices and their stable excited states}

The fundamental SVs and excited states generated from them, with chemical
potential $\mu <0$, can be produced by the following ansatz, written in
polar coordinates $\left( r,\theta \right) $, and adopted as initial
conditions for imaginary-time simulations \cite{ITP1,ITP2}:
\begin{equation}
\phi _{\pm }=A_{\pm }r^{|S_{\pm }|}\exp (-i\mu t -\alpha _{\pm
}r^{2}+iS_{\pm }\theta ),  \label{SVES}
\end{equation}
where $A_{\pm }$ and $\alpha _{\pm }$ are positive constants, and integer
topological-charge numbers, $S_{\pm }$, are related by
\begin{equation}
S_{-}=S_{+}+1  \label{-+}
\end{equation}%
\cite{SVS1}. Fundamental SVs are produced by $(S_{+},S_{-})=(-1,0)$ or $%
(0,+1)$, while the excited state of SVs correspond to $(S_{+},S_{-})=(n,n+1)$%
, where $n\neq -1$ or $0$. In fact, generic solutions in the form of SVs and
their excited states exactly agree with a generalized form of the ansatz
based on Eqs. (\ref{SVES}) and (\ref{-+}) \cite{SVS1}:
\begin{equation}
\begin{aligned} & \phi _{+}=r^{\left\vert S_{+}\right\vert }\Phi _{+}(r)\exp
(iS_{+}\theta ),\\ &\phi _{-}=r^{\left\vert S_{+}+1\right\vert }\Phi
_{-}(r)\exp [i\left( S_{+}+1\right) \theta ], \end{aligned}  \label{ansatz}
\end{equation}%
with real functions taking finite values at $r=0$ and exponentially decaying
$\sim \exp \left( -\sqrt{-\mu }r\right) $ at $r\rightarrow \infty $.

Another type of 2D self-trapped composite states supported by the SO-coupled
BEC corresponds to MMs and their excited states, which and can be obtained
starting from the following ansatz:
\begin{eqnarray}
&&\phi _{\pm }=A_{1}r^{|S_{1}|}\exp (-\alpha _{1}r^{2}\pm iS_{1}\theta )
\notag \\
&&\quad \quad \mp A_{2}r^{|S_{2}|}\exp (-\alpha _{2}r^{2}\mp iS_{2}\theta ),
\label{MMES}
\end{eqnarray}%
where $A_{1,2}$ and $\alpha _{1,2}$ are again positive constants, and
topological-charge numbers are subject to a relation similar to Eq. (\ref{-+}%
): $S_{2}=S_{1}+1$. Actually, ansatz (\ref{MMES}) is the superposition of a
pair of expressions in the form of ansatz (\ref{SVES}) with $S_{+}=S_{1}$
and its mirror image, with $S_{+}=-\left( S_{1}+1\right) $. On the other
hand, unlike the exact ansatz (\ref{ansatz}) for the SVs, there is no exact
generic expression for the MMs. Fundamental MMs correspond to $\left(
S_{1},S_{2}\right) =\left( -1,0\right) $ and $\left( S_{1},S_{2}\right)
=\left( 0,+1\right) $, whereas the excited states of MM are generated by
other integer value of $\left( S_{1},S_{2}\right) $.

Because the SO coupling naturally creates the orbital angular momentum (OAM)
in BEC, we define the normalized OAM of each component, and the total
normalized OAM:%
\begin{equation}
\langle L_{\pm }\rangle ={\frac{\int {d\mathbf{r}\phi _{\pm }^{\ast }\hat{L}%
\phi _{\pm }}}{N_{\pm }}},\quad \langle L\rangle ={\frac{N_{+}\langle
L_{+}\rangle +N_{-}\langle L_{-}\rangle }{N}},  \label{L1L2eq}
\end{equation}%
where $\hat{L}=-i(x\partial _{y}-y\partial _{x})$ is the OAM operator.

\begin{figure*}[tbph]
\subfigure[]{\includegraphics[width=1.5\columnwidth]{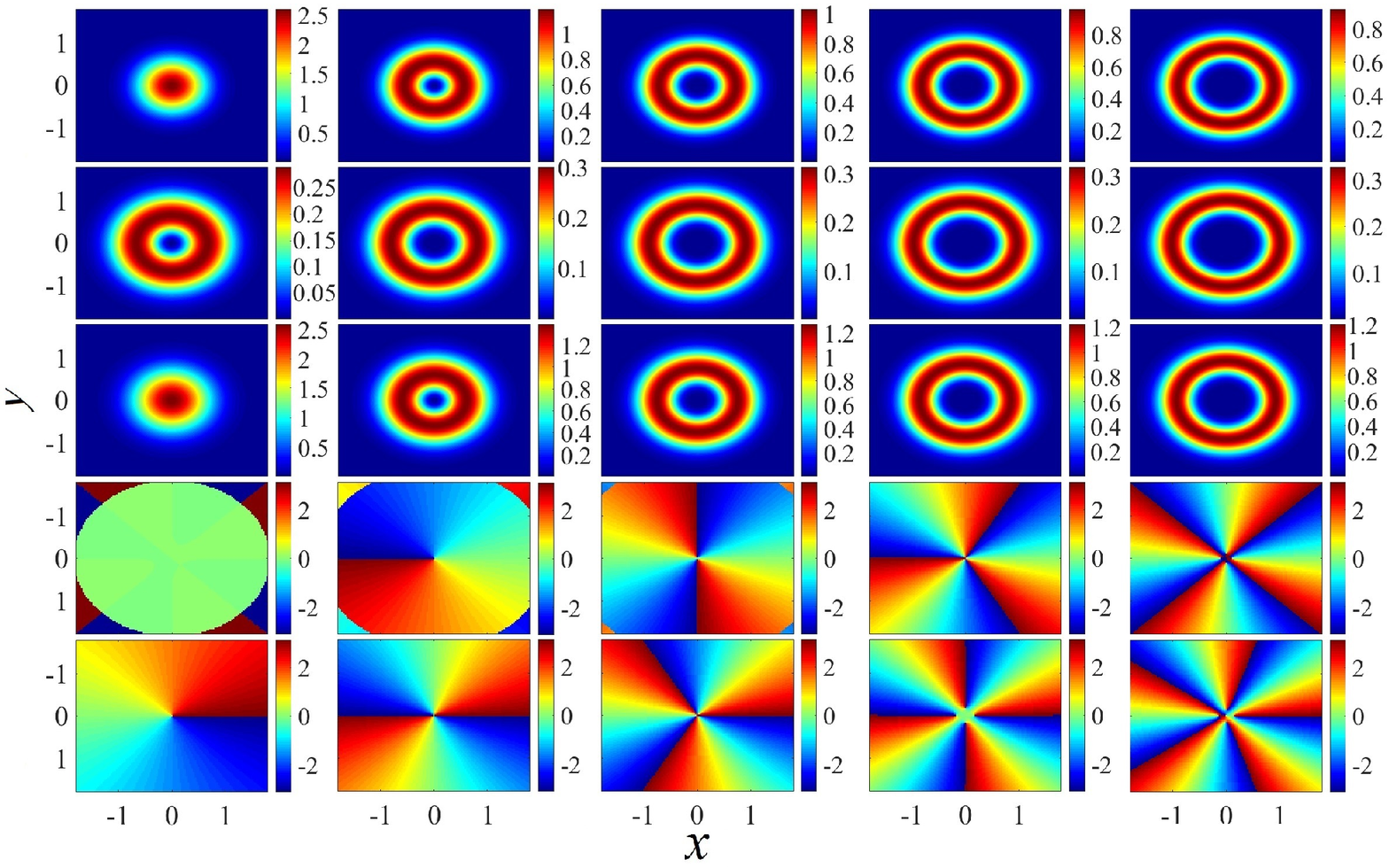}} %
\subfigure[]{\includegraphics[width=1.5\columnwidth]{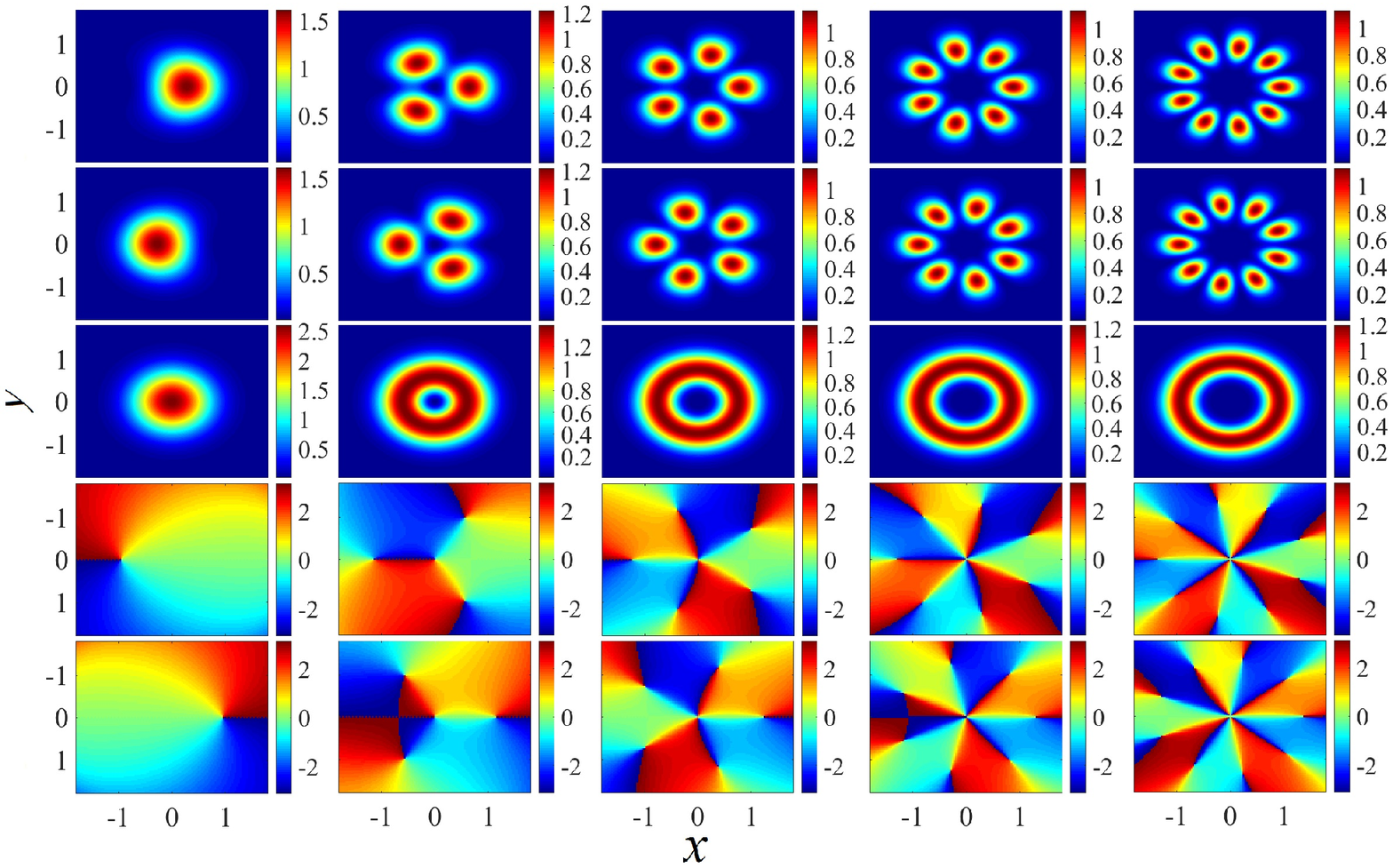}}
\caption{ (Color online) {(a) The first and second rows: density patterns of
the $\protect\phi _{+}$ and $\protect\phi _{-}$ components of the
fundamental SVs and their excited states. The third row: the total-density
patterns of the binary BEC, i.e. $|\protect\phi _{+}|^{2}+|\protect\phi %
_{-}|^{2}$. The fourth and fifth rows: phase patterns of $\protect\phi _{+}$
and $\protect\phi _{-}$, respectively. From left to right: $S_{+}=0,1,2,3$,
and $4$. (b) Similar results for the MMs and their excited states, with the
same meaning of the rows as in (a). From left to right: $S_{1}=0$, $1,2,3$,
and $4$. The total norm of all the soliton modes displayed in the figure is $%
N=4$.} }
\label{MFV1}
\end{figure*}

Stationary solutions for SVs and their excited states were generated by
means of the imaginary-time integration of Eq. (\ref{fulleq}) (where we
adopt normalization $\lambda =1$, as said above), initiated by\ the input
given by Eqs. (\ref{SVES}) and (\ref{-+}) with integer $S_{+}$. Then, the
stability of the so obtained states was tested by means of real-time
simulations. This scenario is relevant for making it sure that the
imaginary-time evolution has produced the relevant solutions (for given
norm). Indeed, it may happen, in some cases, that the trajectory of the
imaginary-time evolution passes very close to a saddle point, and gets stuck
there, which is an obstacle for producing the target states. For this
reason, it is commonly adopted to run real-time simulations following the
imaginary-time integration (see, e.g., Ref. \cite{Kumar2016}).

We apply rescaling to fix $\alpha =1$ in Eq. (\ref{g0}), and,
following Ref. \cite{Yongyao2013}, we take $g_{0}=0$ and $\epsilon =0.5$,
other values of these parameters producing quite similar results. The
imaginary-time simulations readily produce SVs and their excited states with
all integer values of the topological charge, $S_{+}=0,\pm 1,\pm 2,\pm
3,\ldots $. Typical examples of \emph{stable} SVs and their excited states
with $S_{+}=0,1,2,3$, and $4$ are displayed in Fig.\ref{MFV1}(a), the
respective total-density pattern, $|\psi _{+}|^{2}+|\psi _{-}|^{2}$, being a
perfect ring, as seen in the third row of Fig.\ref{MFV1}(a). Further,
families of SVs and their excited states are characterized, as usual, by
dependences $\mu (N)$ and $E(N)$ for them for different values of $S_{+}$,
which are displayed in Fig.\ref{figmuEN} panels (a) and (b). They former
one satisfies the \textit{anti-Vakhitov-Kolokolov }criterion, i.e., $d\mu
/dN>0$, which is a known necessary condition of the stability of bright
solitons supported by self-repulsive nonlinearities \cite{VK1,antiVK}. The
solitons with topological numbers $(S_{+},S_{-})=(n,n+1)$ and $(-n-1,-n)$
(where $n\geq 0$ is an arbitrary integer) obviously have identical energies
and chemical potentials, therefore, we only consider the case of $S_{+}\geq
0 $ for this type of the solitons. In the case of $S_{+}\geq 0$, at $%
N\rightarrow 0$, evident limit values are $\mu \rightarrow |S_{+}|-0.5$ and $%
E\rightarrow 0$ \cite{SVS1}.

The normalized OAM of each component of the SVs and their excited states,
defined by Eq. (\ref{L1L2eq}), are $\langle L_{\pm }\rangle \equiv S_{\pm }$%
, i.e., both components of the excited states of SVs are eigenstates of the
OAM operator. The component with a smaller value of $|S_{\pm }|$ accounts
for a larger contribution to the total norm, in agreement with the trend
found for the SVs in Ref. \cite{SVS1}. The total normalized OAM, $\langle
L\rangle $, is displayed vs. $N$ for different values of $S_{+}$ in Fig.\ref%
{figmuEN}(c), with $\langle L\rangle =S_{+}+0.5$ in the limit of $%
N\rightarrow 0$. At $N>1$, $\langle L\rangle $, which is slightly smaller
than $S_{+}+0.5$, almost does not depend on $N$. The latter result indicates
that the SVs and their excited states may be approximately regarded as
effectively having a half-integer eigenvalue of the normalized OAM.

As mentioned above, the stability of the SVs and their excited states was
verified by means of direct real-time simulations, using the split-step
fast-Fourier-transform algorithm. We have thus found that the excited state
of SVs are completely stable in this setting for $S_{+}\leq 5$. A typical
example of the stable evolution of the excited state with $S_{+}=5$ is
displayed in Fig.\ref{figRTP}(a). At $S_{+}\geq 6$, the excited states of
SVs are less stable, with some corrugation appearing in the vortex rings in
the course of the long-time propagation. Nevertheless, the excited state of
SVs keep their overall vortex structure. A typical example of them with $%
S_{+}\geq 6$ is displayed in Fig.\ref{figRTP}(b).

The observed stabilization of the vortex solitons with large values of $S$
is a noteworthy findings, as such solitons tend to be unstable in the
majority of models of nonlinear media, spontaneously splitting into
fragments the number of which being equal or close to $|S|$ \cite%
{SotoCrespo1991,Skryabin1998,Sears1998,Desyanikow2001}. Stable vortex
solitons with $S>1$ have \ been found in Bessel lattices with
self-defocusing nonlinearity \cite{Kartashov2005} and, more recently, in
radial ring-lattices with repulsive dipole-dipole interactions \cite%
{Huang2017}, as well as in the other model with the effectively nonlinear
interaction, which corresponds to a binary BEC with the components coupled
by a microwave field \cite{Dong} (in the latter case, the components
represent two hyperfine atomic states coupled by a magnetic transition). In
this connection, it is relevant to mention that matter-wave fields carrying
definite values of the OAM have various applications to quantum-information
processing and optical communications \cite%
{Allen1992,Franke-Arnold2008,Yusiyuan2012,JWang2012,Padgett2015,Ding2015,Miao2016}%
, hence stable excited state of SVs with large values of $S$ may help to
expand the range of the potential applications \cite{Dong}.

\section{Mixed modes and their stable excited states}

Stationary solutions for the MMs and their excited states were produced by
imaginary-time simulations of Eq. (\ref{fulleq}) with input (\ref{MMES}),
setting $S_{1}=0,\pm 1,\pm 2,\ldots $. The solutions produced by $%
S_{1}=-(n+1)$ are tantamount to those obtained with $S_{1}=n$ ($n\geq 0$),
therefore, we consider only the case of $S_{1}\geq 0$ for states of this
type. Unlike the SVs and their excited states, the modes of the MMs and
their excited states types naturally have equal norms of the two components,
i.e., $N_{+}=N_{-}$. Typical examples of stable MMs and their excited states
with $S_{1}=0$ ,$1$, $2$, $3$ and $4$ are displayed in Fig.\ref{MFV1}(b).
Numerical results reveal that the established excited states of MM in each
component is built as a necklace ring, with the number of fragments exactly
equal to $S_{1}+S_{2}\equiv 2S_{1}+1$.

\begin{figure}[tbph]
\subfigure[]{\includegraphics[width=0.48\columnwidth]{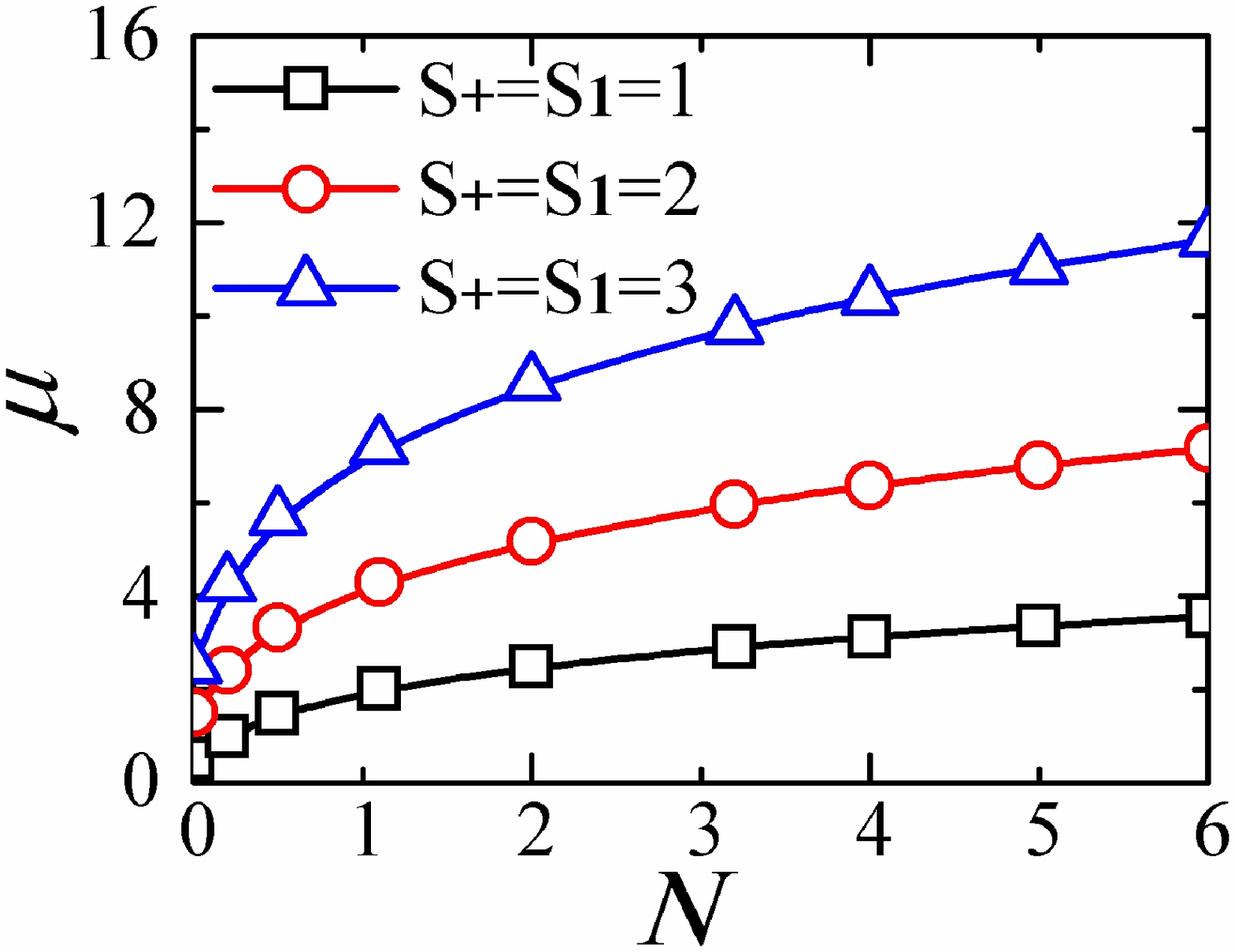}} %
\subfigure[]{\includegraphics[width=0.48\columnwidth]{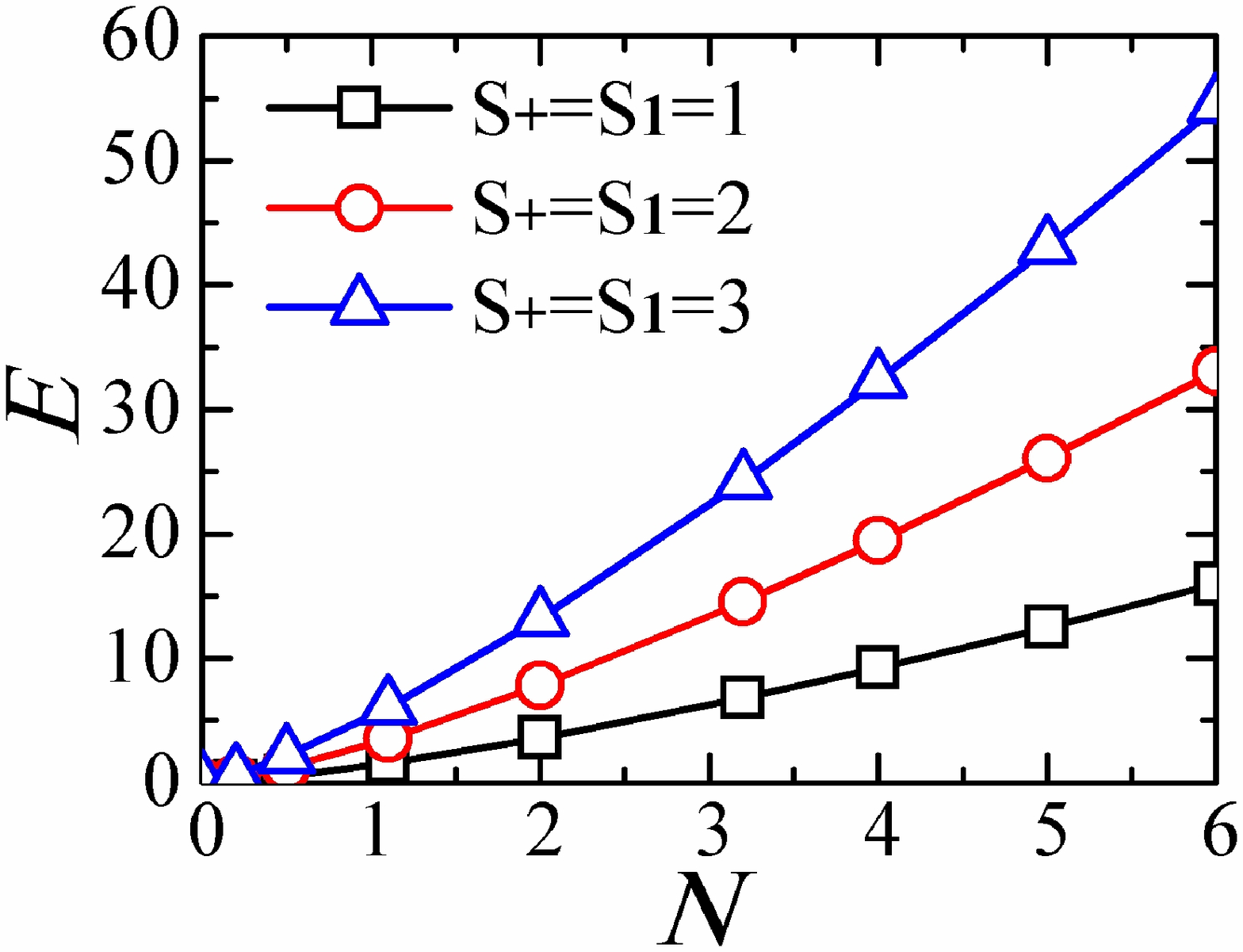}} %
\subfigure[]{\includegraphics[width=0.48\columnwidth]{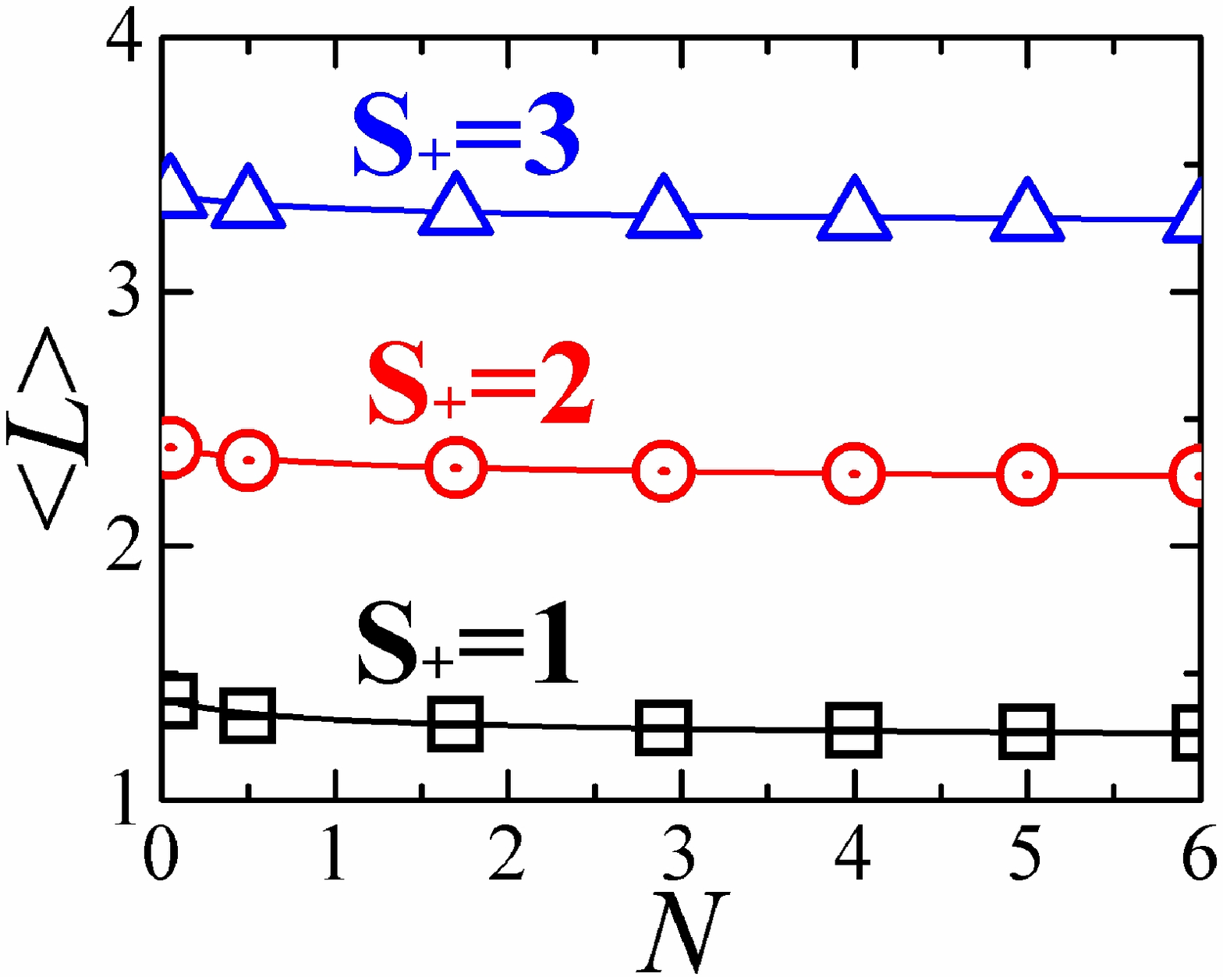}} %
\subfigure[]{\includegraphics[width=0.48\columnwidth]{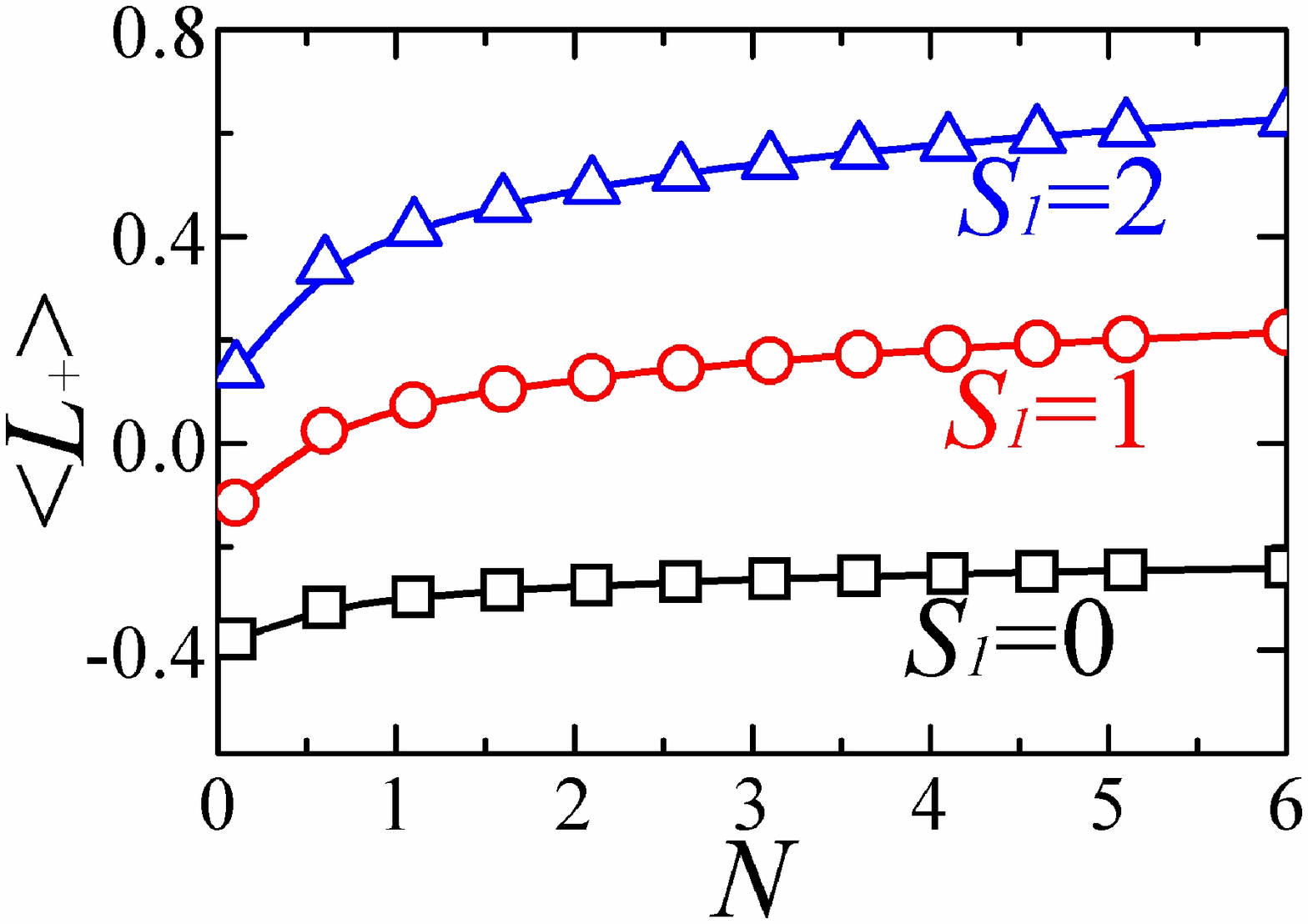}} %
\subfigure[]{\includegraphics[width=0.48\columnwidth]{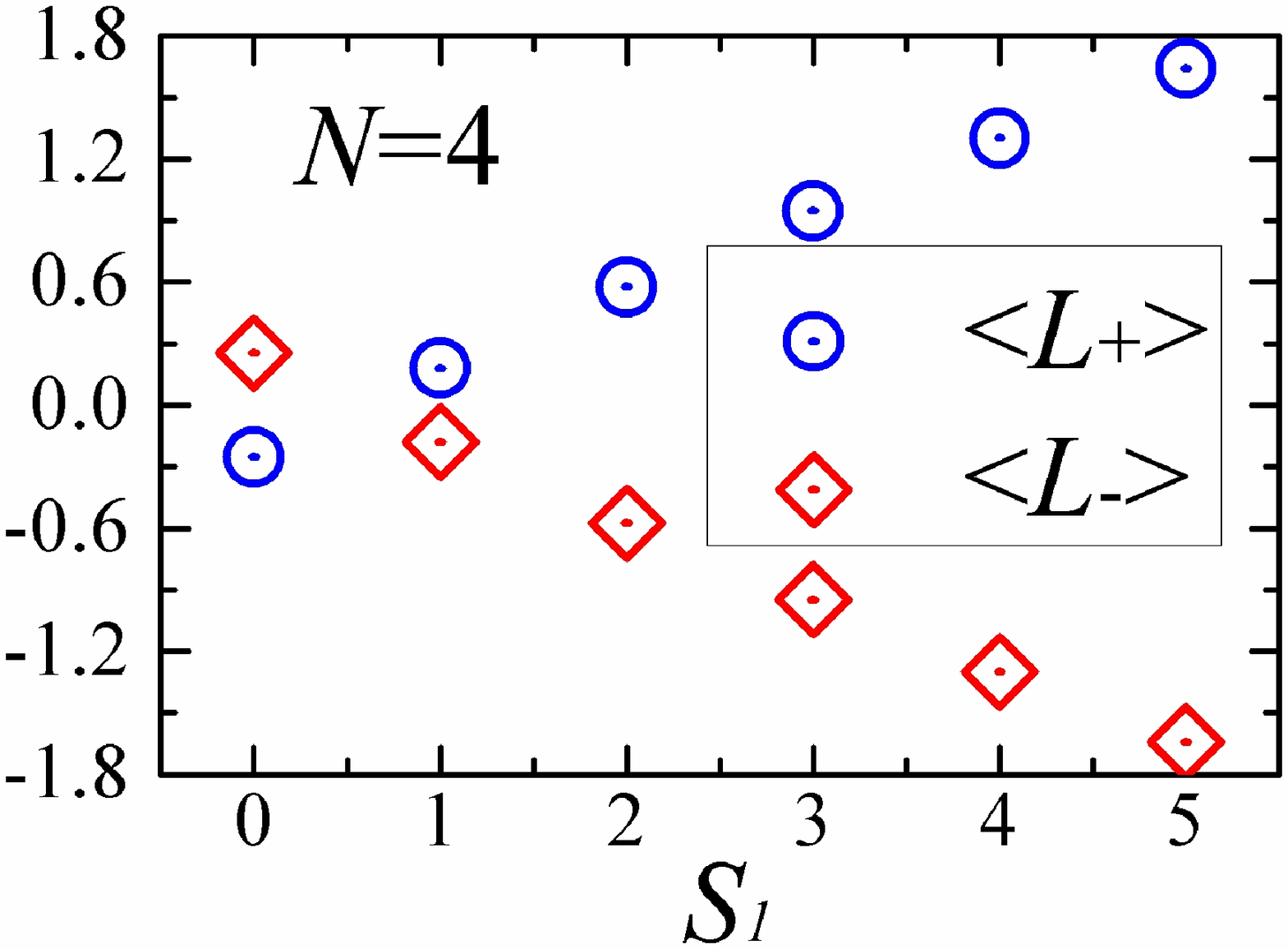}}
\caption{(Color online) {(a) The chemical potential, $\protect\mu $, and (b)
energy, $E$, for \emph{both} excited states of SVs and MM as functions of
the total norm, $N $ (the coincidence of the curves for the excited states
of SVs and MM is explained in the main text). Black curves with squares, red
curves with circles, and blue curves with triangles correspond to $
S_{+}=S_{1}=1$, $2$, and $3$, respectively. (c) The total normalized orbital
angular momentum values $\langle L\rangle $ of excited state of SVs versus $%
N $ for different values of $S_{+}$; (d) the normalized orbital angular
momentum for component $\protect\phi _{+}$, i.e., $\langle L_{+}\rangle $,
of excited states of MM versus $N$ for different values of $S_{1}$; (e)
values of $\langle L_{\pm }\rangle $ for excited states of MM versus $S_{1}$
for $N=4$. (note that $\langle L_{-}\rangle \equiv -\langle L_{+}\rangle $
for excited states of MM).} }
\label{figmuEN}
\end{figure}

\begin{figure}[tph]
%[width=0.3\columnwidth]
\subfigure[]{\includegraphics[width=0.48\columnwidth]{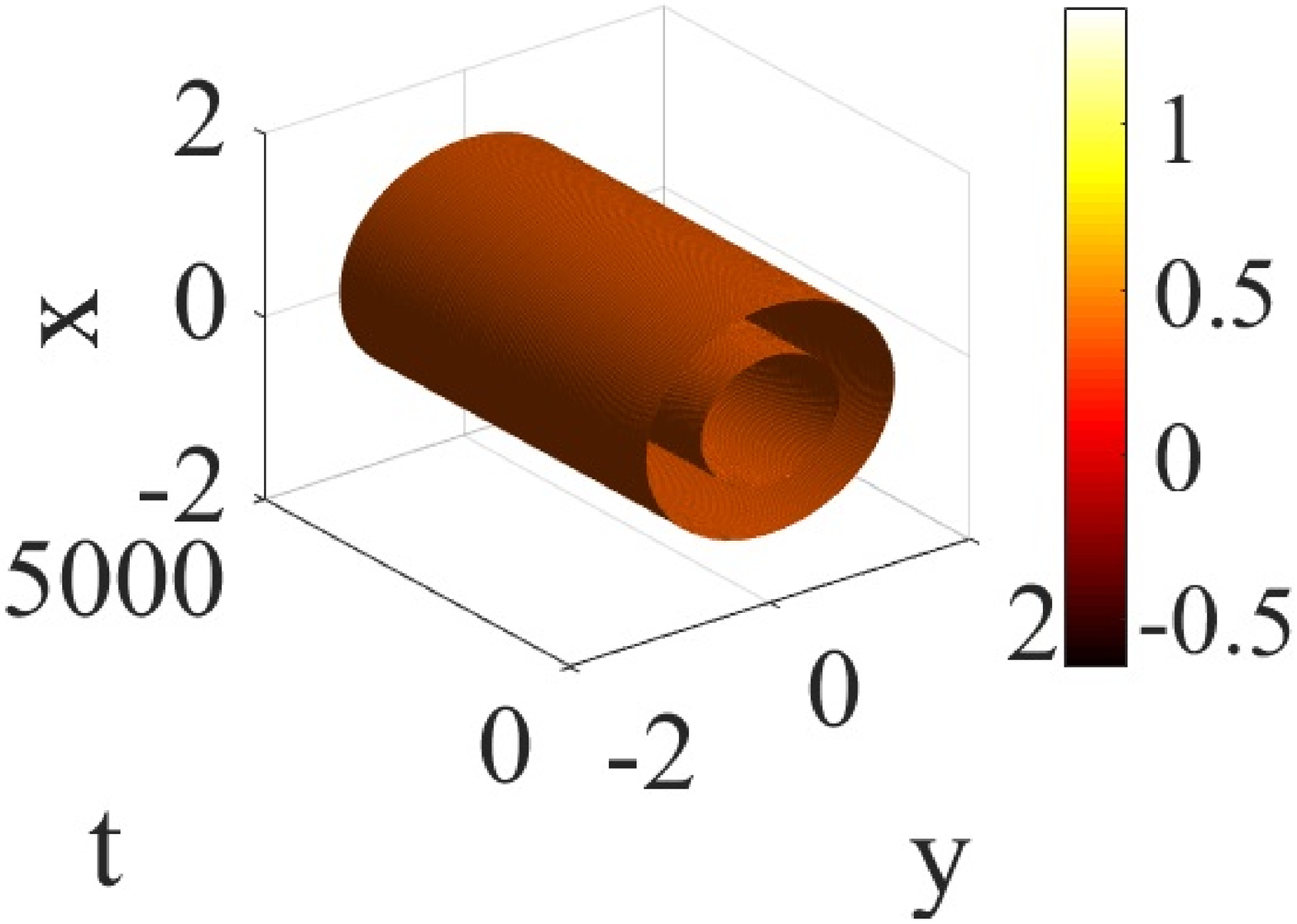}} %
\subfigure[]{\includegraphics[width=0.48\columnwidth]{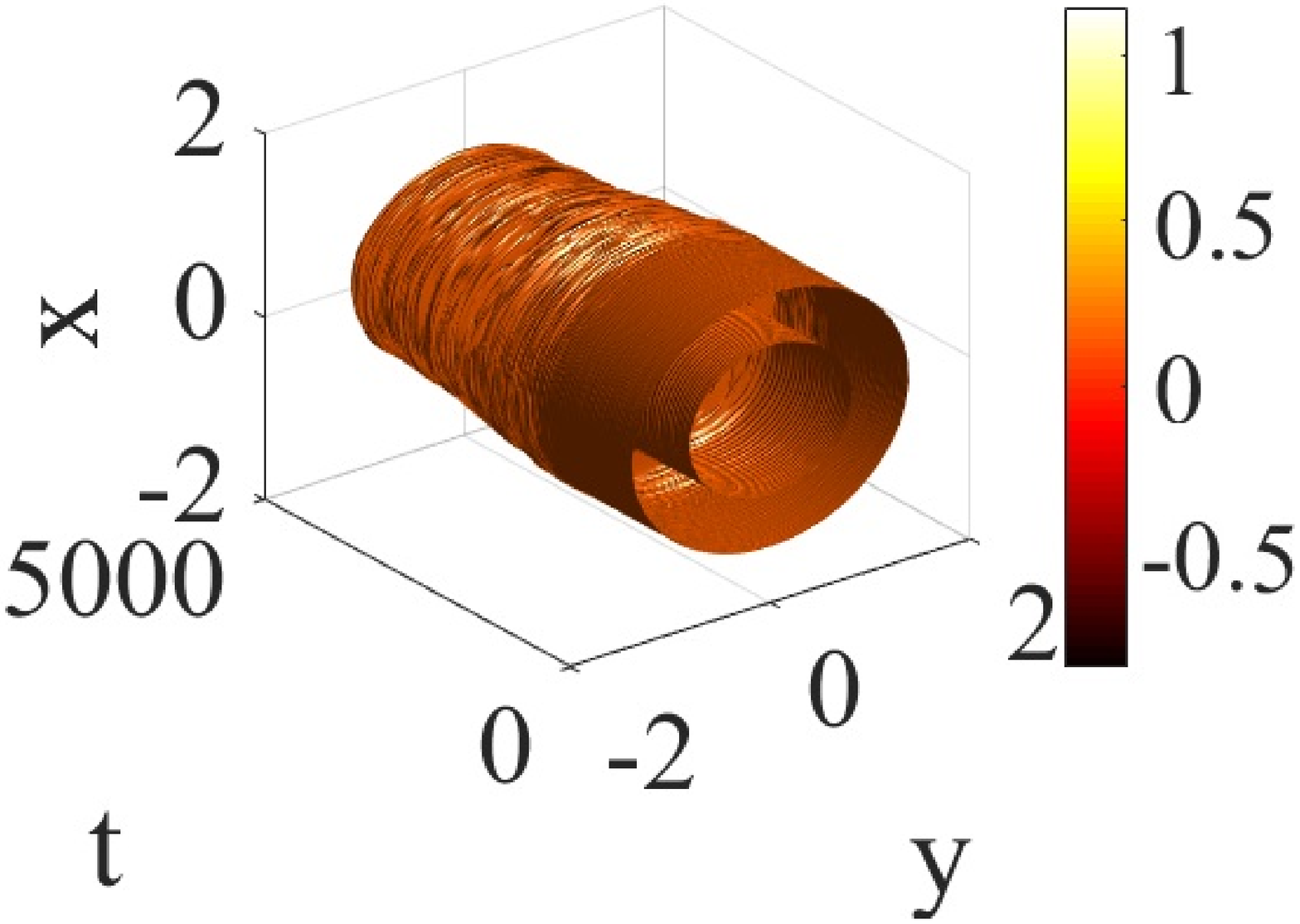}} %
\subfigure[]{\includegraphics[width=0.48\columnwidth]{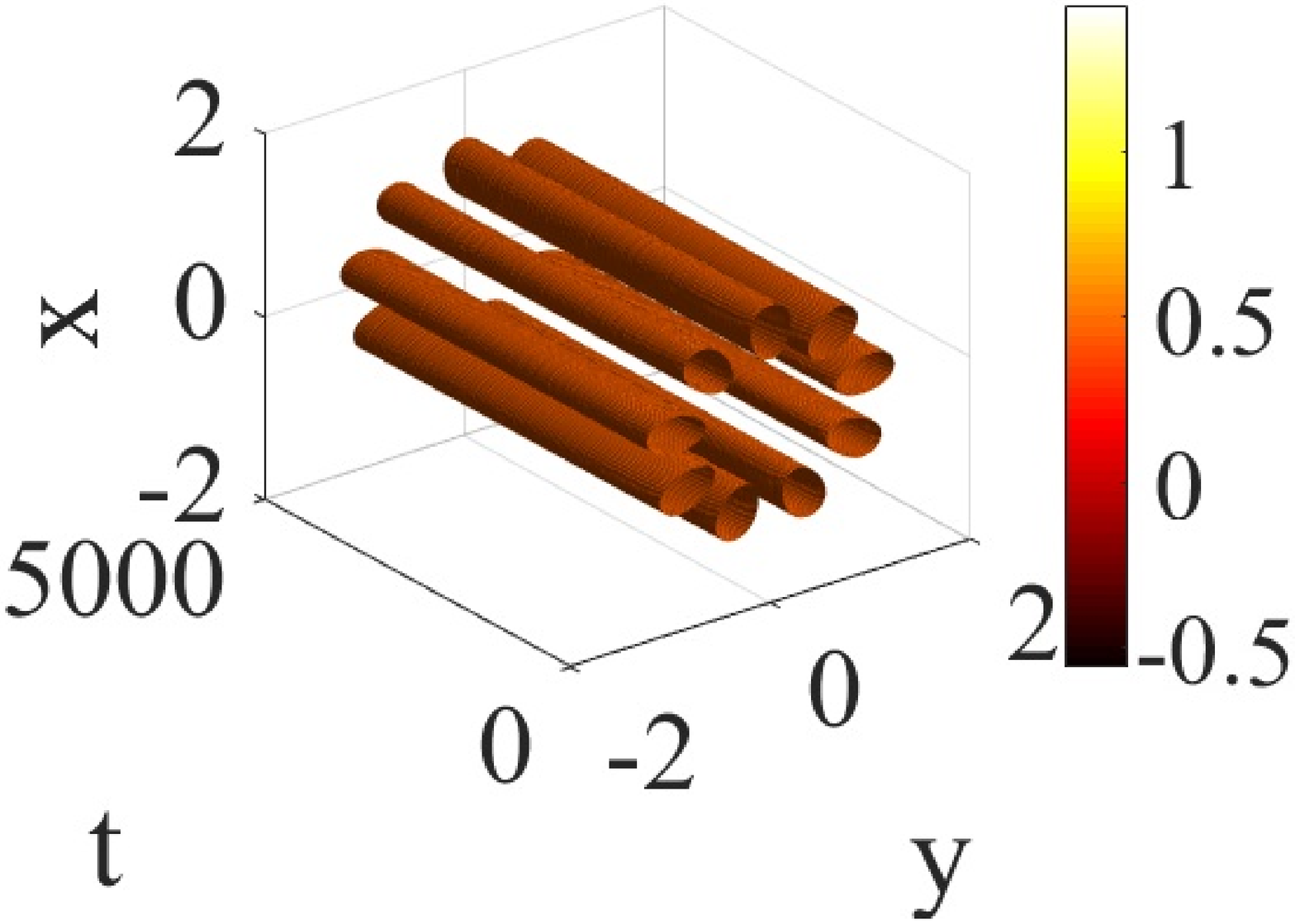}} %
\subfigure[]{\includegraphics[width=0.48\columnwidth]{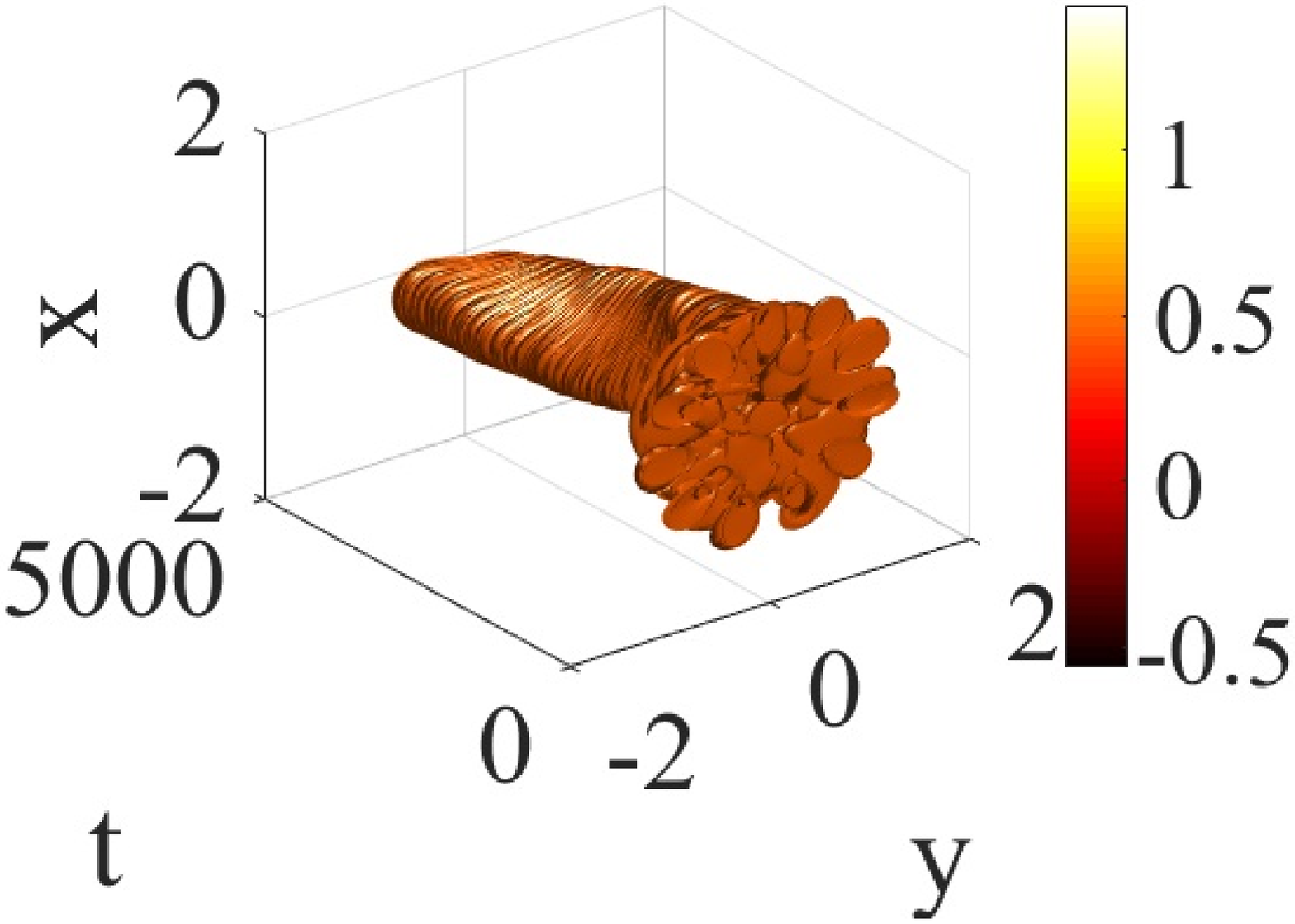}}
\caption{(Color online) Long real-time evolution of the density pattern $|%
\protect\psi _{+}(t)|^{2}$, with $3\%$ random noise added to the initial
conditions in all panels here. (a) A stable excited state of SVs mode with $%
(N,S_{+})=(4,5)$. (b) An unstable excited state of SVs with $(N,S_{+})=(4,6)$%
. (c) A stable excited state of MM with $(N,S_{1})=(4,4)$. (d) An unstable
excited state of MM with $(N,S_{1})=(4,5)$. All the panels are shown as
isosurfaces of $0.4\times |\protect\psi _{+}^{\max }(t=0)|^{2}$. }
\label{figRTP}
\end{figure}

The results reported in Ref. \cite{SVS1} demonstrated that chemical
potentials and energies of the SVs and MMs coincide ($\mu _{\mathrm{SV}}=\mu
_{\mathrm{MM}}$ and $E_{\mathrm{SV}}=E_{\mathrm{MM}}$) for the systems of
the Manakov's type, with equal coefficients of the contact self-attraction
and cross-attraction \cite{Manakov}. Because dipole-dipole interactions in
Eq. (\ref{fulleq}) automatically satisfy the Manakov's condition, the
excited states of SVs and MM with $S_{+}=S_{1}$ also have equal chemical
potentials and energies.\ Therefore, the plots in Fig.\ref{figmuEN}(a) and
(b) represent curves of $\mu (N)$ and $E(N)$, respectively, for excited
states of SVs and MM alike, for given values of $S_{1}$. The stability of
the excited states was verified through direct real-time simulations. We
have found that excited states of MM are stable at $S_{1}\leq 4$ and
unstable at $S_{1}\geq 5$. Typical examples of the evolution of stable and
unstable excited states of MM are displayed in Fig.\ref{figRTP}(c,d). In
particular, Fig.\ref{figRTP}(d) shows that unstable excited states of MM
develop spontaneous twist at the initial stage of the evolution, and
eventually degenerate into fundamental MMs.

The complexity of the phase patterns of the excited states of MM increases
with the increase of the number of fragments in the corresponding necklace,
i.e., with the growth of $S$. The ansatz written in Eq. (\ref{MMES})
demonstrates that the MMs and their excited states may be considered as
mixtures of two different eigenstates of the OAM\ operator, therefore the
normalized OAM of each component, defined as per\ Eq. (\ref{L1L2eq}), no
longer takes integer values, being functions of $N$ and $S_{1}$, with the
total normalized OAM\ being zero, $\langle L_{+}\rangle +\langle
L_{-}\rangle \equiv 0$. Figures \ref{figmuEN}(d) and (e) show $\langle
L_{+}(N)\rangle $ for different values of $S_{1}$, and $\langle L_{\pm
}(S_{1})\rangle $ for $N=4$, respectively. These figures demonstrate that $%
\langle L_{+}\rangle $ gradually increases with the growth of $N$ and $S_{1}$%
.

It is interesting to note that spatial positions of fragments
(\textquotedblleft beads") of the necklace in one component are located
exactly in the middle of positions of two adjacent fragments of the other
component [see the panels in first and second rows of Fig.\ref{MFV1}(b)].
This feature causes the overall density pattern, $|\phi _{+}|^{2}+|\phi
_{-}|^{2}$, to exhibit a profile of a perfect ring. For a given value of
total norm $N$, the total-density patterns of the excited states of SVs and
MM are identical for $S_{1}=S_{+}$ [see the third rows in Fig.\ref{MFV1}%
(a,b)], while, as mentioned above, the total normalized OAM of the excited
states of MM vanishes, $\langle L\rangle =(\langle L_{+}\rangle +\langle
L_{-}\rangle )/2\equiv 0$, on the contrary to nonzero $\langle L\rangle $
for the excited states of SVs. Moreover, because the normalized OAM for each
component is different from zero, $\langle L_{+}\rangle =-\langle
L_{-}\rangle \neq 0$, the two-component MMs and their excited states
resemble, in terms of photonics, linearly polarized light split into left
and right circularly polarized components. In this connection, it is
relevant to mention that the splitting of linearly polarized light beams
into left- and right-polarized waves finds diverse applications in chiral
media and topological photonics \cite{Mciver2012,YHWang2013,Bliokh2015,Guo2015,Rafayelyan2016,Lodahl2017}, suggesting that similar
applications, such as the realization of topological insulators, may be also
realized in terms of the matter waves.

\section{Conclusion}

The objective of this work was to stabilize the 2D excited states of SVs
(semi-vortices) and MM (mixed mode) in spin-orbit-coupled BECs in the
setting based on the dipole-dipole interactions between originally isotropic
atoms, induced by a polarizing field oriented perpendicular to the plane in
which the BEC is trapped, under the assumption that the strength of the
polarizing field grows in the radial direction as $r^{4}$. Stable excited
states of SVs and MM are predicted in this setting for the first time. They
have a size $\sim 10$ $\mathrm{\mu }$m, and may contain up to $10^{5}$
atoms. Both components of the excited states of SVs are eigenstates of the
OAM (orbital angular momentum), the total normalized OAM of such a soliton
being $\langle L\rangle \approx S_{+}-0.5$. Characteristics and stability of
these excited states of SVs have been systematically studied. They are
completely stable at $S_{+}\leq 5$, and become weakly unstable for $%
S_{+}\geq 6$. The excited states of MM feature a circular necklace structure
with mutually interleaved components, whose total-density pattern is a
perfect ring. The values of the normalized OAM of the two components of the
excited states of MM are $\langle L_{+}\rangle =-\langle L_{-}\rangle $,
with the vanishing total OAM, $\langle L_{+}\rangle +\langle L_{-}\rangle
\equiv 0$. The excited states of MM are stable at $S_{1}\leq 4$, and become
unstable at $S_{1}\geq 5$. The characteristics of the stable excited states
of SVs and MM suggest that they may find potential applications to
high-precision communications, as well as to the design of chiral media and
topological insulators for matter waves.

\begin{acknowledgments}
C. Huang and Y. Li appreciate the very useful discussions from Prof.
Chaohong Lee. This work is supported, in a part, by the National Natural
Science Foundation of China (Grant Nos. 11575063, 61705035, 11204037,
61575041), and by Grant No. 2015616 from the joint program in physics
between the NSF (US) and Binational (US-Israel) Science Foundation, and by
Grant No. 1286/17 from the Israel Science Foundation.
\end{acknowledgments}

\bibliographystyle{plain}
\bibliography{apssamp}
% Produces the bibliography via BibTeX.

\end{document}